\documentclass[aps,prx,reprint,amsmath,amssymb,twocolumn, superscriptaddress, floatfix]{revtex4}
\usepackage{tikz}
\usepackage{graphicx}
\usepackage{textcomp}
\newcommand{\ket}[1]{|#1\rangle}

\usepackage{hyperref, soul}
\usepackage[capitalize]{cleveref}
\hypersetup{
    colorlinks,
    linkcolor={blue},
    citecolor={blue},
    urlcolor={blue}
}

\usepackage{xcolor}

\definecolor{purple_plot}{RGB}{97,64,155}
\definecolor{green_plot}{RGB}{92,185,107}

\begin{document}

\title{Characterising the failure mechanisms of error-corrected quantum logic gates}

\author{Robin Harper}
\affiliation{Centre for Engineered Quantum Systems, School of Physics, The University of Sydney, Sydney, New South Wales 2006, Australia}

\author{Constance Lain\'{e}}
\affiliation{Centre for Engineered Quantum Systems, School of Physics, The University of Sydney, Sydney, New South Wales 2006, Australia}
\affiliation{London Centre for Nanotechnology, University College London, London WC1H 0AH, United Kingdom}

\author{Evan Hockings}
\affiliation{Centre for Engineered Quantum Systems, School of Physics, The University of Sydney, Sydney, New South Wales 2006, Australia}

\author{Campbell McLauchlan}
\affiliation{Centre for Engineered Quantum Systems, School of Physics, The University of Sydney, Sydney, New South Wales 2006, Australia}

\author{Georgia M. Nixon}
\affiliation{Centre for Engineered Quantum Systems, School of Physics, The University of Sydney, Sydney, New South Wales 2006, Australia}

\author{Benjamin J. Brown}
\affiliation{IBM Quantum, T. J. Watson Research Center, Yorktown Heights, New York 10598, USA}
\affiliation{IBM Denmark, Sundkrogsgade 11, 2100 Copenhagen, Denmark}

\author{Stephen D. Bartlett}
\email{stephen.bartlett@sydney.edu.au}
\affiliation{Centre for Engineered Quantum Systems, School of Physics, The University of Sydney, Sydney, New South Wales 2006, Australia}

\begin{abstract}
Mid-circuit measurements used in quantum error correction are essential in quantum computer architecture, as they read out syndrome data and drive logic gates. Here, we use a heavy-hex code prepared on a superconducting qubit array to investigate how different noise sources impact error-corrected logic.
First, we identify that idling errors occurring during readout periods are highly detrimental to a quantum memory. We demonstrate significant improvements to the memory by designing and implementing a low-depth syndrome extraction circuit. Second, we perform a stability experiment to investigate the type of failures that can occur during logic gates due to readout assignment errors.
We find that the error rate of the stability experiment improves with additional stabilizer readout cycles, revealing a trade-off as additional stability comes at the expense of time over which the memory can decay. We corroborate our results using holistic device benchmarking and by comparison to numerical simulations. Finally, by varying different parameters in our simulations we identify the key noise sources that impact the fidelity of fault-tolerant logic gates, with measurement noise playing a dominant role in logical gate performance.
\end{abstract}

\maketitle

\section{Introduction}

The logical operations of an error-corrected quantum processor are driven by the outcomes of measurements performed throughout the execution of a quantum circuit. 
Not only are these mid-circuit measurements used as a syndrome for errors acting on logical qubits stored in memory \cite{sundaresan2022, ryan-anderson2021, Andersen-Wallraff-RepeatedQuantumError-2020, google24, Postler2024}, but the outcomes of these measurements are also used to implement logical gates such as lattice surgery operations \cite{Bombin_2009, fowler2012b, Horsman2012,Gupta2024,RyanAnderson2024,lacroix2024scalinglogiccolorcode,besedin2025realizinglatticesurgerydistancethree, Raussendorf2007, Brown2017, Litinski2019gameofsurfacecodes, Cohen2022,Thomsen2022,Davydova2025}. 
To deal with the fact that errors may give rise to unreliable measurement outcomes that could potentially lead to a corrupted logical gate, we design logic gates to be fault tolerant by adding redundancy, such that the error syndrome can be used to identify errors in the measurements themselves.

A straightforward way to include such redundancy is to repeat the measurements we need to complete a logic gate multiple times~\cite{Dennis2002}.  
Increasing the number of repetitions of mid-circuit measurements can decrease the probability of a logical gate failure. However, this leads to a trade-off since, by increasing the number of rounds of measurements in a logic gate, we increase the period over which a memory must store logical quantum information, and as such there is an increased chance of logical corruption. Ideally, we should optimise our logical operations to minimise the probability of both memory corruption together with logic gate failure with respect to the underlying hardware. Identifying the bottlenecks under this optimisation over logical error rates will show us new pathways to improve hardware towards the development of a fault-tolerant quantum computer.

Here we demonstrate this trade-off through experiments performed by preparing the heavy-hex code on a superconducting quantum processor~\cite{chamberland2020}. 
First, we conduct a memory experiment~\cite{Andersen-Wallraff-RepeatedQuantumError-2020,GoogleQuantumAI-ExponentialSuppressionBit-2021,ryan-anderson2021,zhao2022,Krinner2022} to quantify the logical error rate for different numbers of syndrome measurement rounds. 
We significantly improve this logical error rate for the heavy-hex code by decreasing the circuit depth and real-time duration of the stabilizer readout circuit with two innovations.  Our first innovation is a new circuit to learn the syndrome data that has significantly smaller circuit depth compared with previous implementations of this code. Our new circuit is able to defer the readout of one type of check such that both the Pauli-$X$ and Pauli-$Z$ type checks are measured in parallel, thereby decreasing the number of sequential rounds of measurements we need to perform to complete a full syndrome readout cycle. Given that mid-circuit measurement times dominate the syndrome extraction circuit cycle, we find this leads to a substantial speedup and corresponding reduction in logical error rate. Our second innovation is to replace the reset operation that follows a measurement with a classical Pauli frame update \cite{geher2024}, eliminating the need for reset of ancilla qubits, and yielding a significant speed up and further reduction in the logical error rate. Altogether these improvements result in a logical qubit encoded in a heavy-hex code with a survival probability of $96\%$ per round of syndrome extraction.

We complement our memory experiment with a new stability experiment~\cite{Gidney2022dual,caune_demonstrating_2024} designed for the heavy-hex code.
A stability experiment benchmarks the performance of a fault-tolerant logic gate implemented with lattice surgery. A logical error in stability is incurred if an unfortunately-located sequence of mid-circuit measurement failures occurs.
We measure the logical gate failure rate as a function of measurement rounds.
We identify a decay in logical error rate as a function of the number of measurement rounds, indicating below-threshold behaviour.
We compare experiments where we use resets to experiments where we replace post-measurement resets with a Pauli frame update.
Notably, in contrast to the error models and simulations used in Ref.\ \cite{geher2024}, we find very little difference in the two experiments. We attribute this to the reset mechanism used in the quantum device. We support this assertion with simulations.

We corroborate our error correction results with supporting benchmarking experiments and numerical simulations. We use benchmarking circuits to learn the noise the device experiences during syndrome readout circuits with mid-circuit measurements to determine the extent to which the device respects a circuit noise model. Our numerical simulations show the dependence of the different parameters, 
on physical device parameters assuming a circuit model, e.g., gate error rate, measurement error rate, and idling errors.

Our experiments and our supporting analysis allow us to critique the performance of current quantum hardware, and how their development should progress in order for future generations of devices to be able to perform large-scale fault-tolerant logic operations. 
Investigating this trade-off reveals that the logical error rate of a fault-tolerant logic gate in our studied quantum processor is dominated by the logical gate failure rate, rather than the logical memory failure rate. 
Given that measurement error rates are a dominant contributor to the value of sub-threshold performance for all types of logical failures, our results indicate that, currently, better mid-circuit measurements, in terms of both their error rates, and measurement times, will significantly improve the performance of logic gates. 

\section{Results}

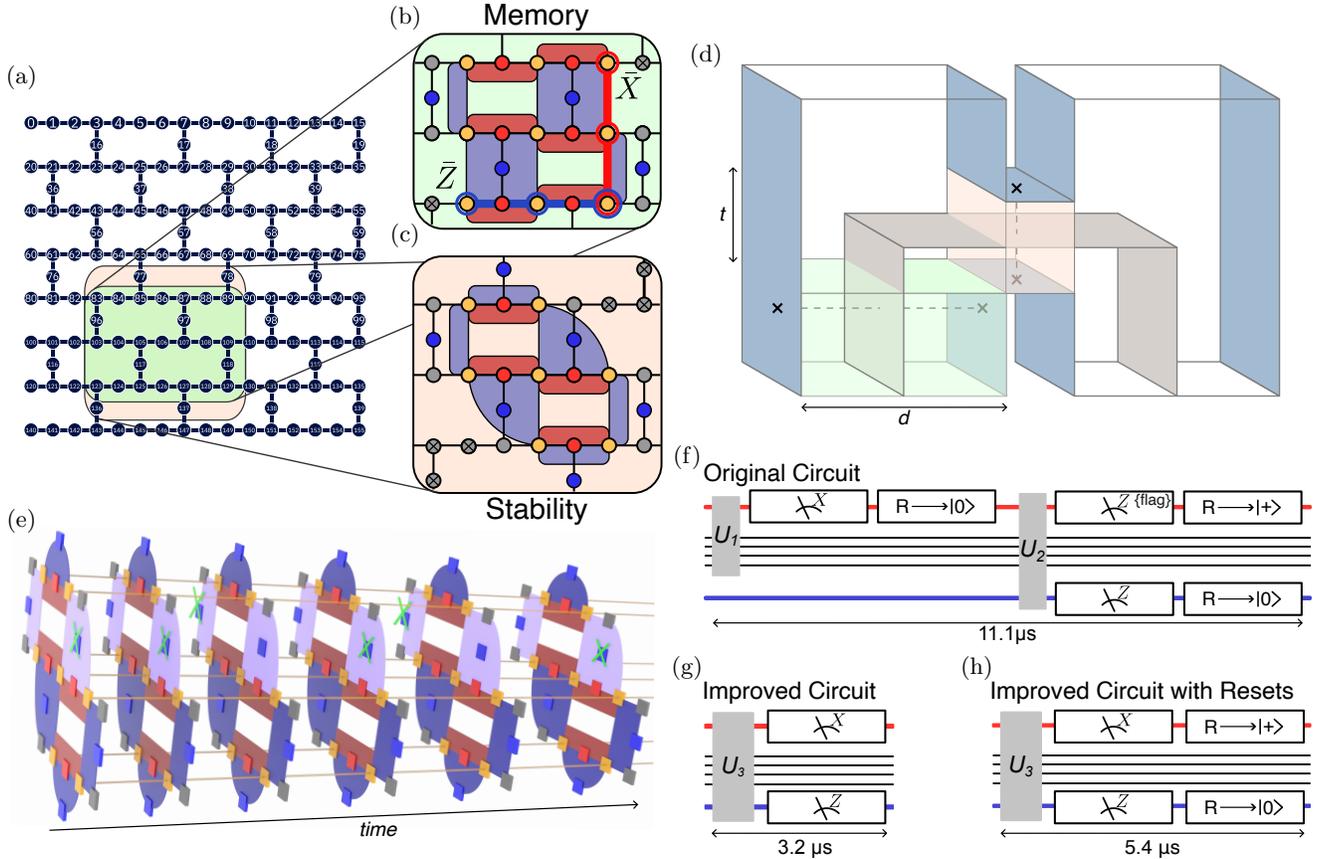
\begin{figure*}
\begin{tikzpicture}
    \node[inner sep=0pt] at (0,0) {\includegraphics{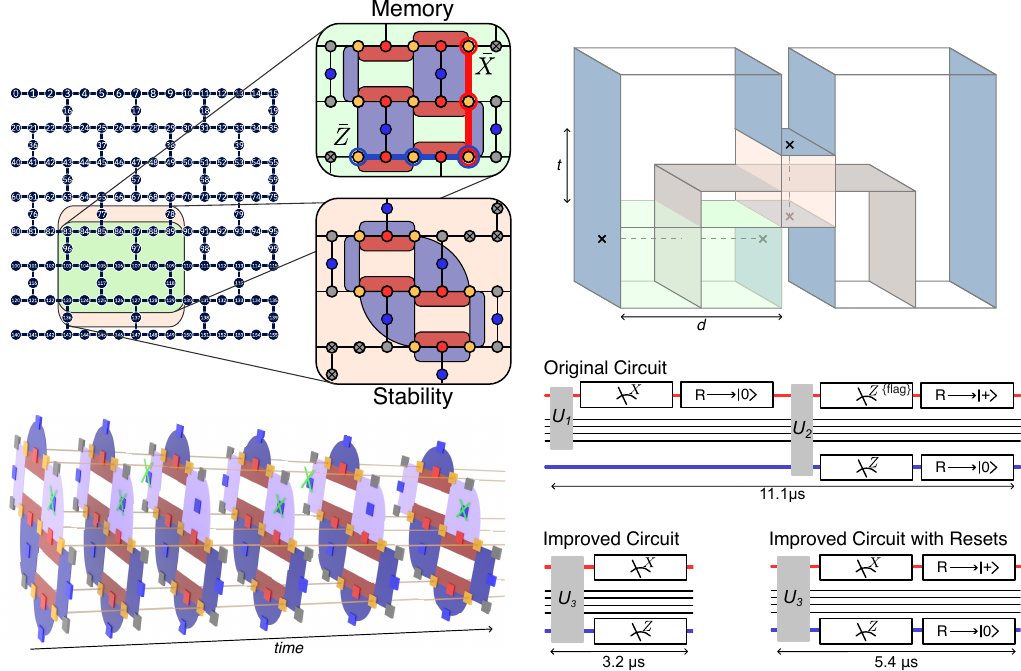}};
\node[inner sep=0pt] at (-8.5,4.7)         {(a)};
\node[inner sep=0pt] at (-3.4,5.5)         {(b)};
\node[inner sep=0pt] at (-3.4,2.6)         {(c)};
\node[inner sep=0pt] at (0.6,5)         {(d)};
\node[inner sep=0pt] at (-8.5,-1.2)         {(e)};
\node[inner sep=0pt] at (0.35,-0.35)         {(f)};
\node[inner sep=0pt] at (0.35,-3.15)         {(g)};
\node[inner sep=0pt] at (4.2,-3.15)         {(h)};
\end{tikzpicture}
    \caption{\textbf{Memory and stability circuits.} 
    \textbf{(a)} Heavy-hex layout of the IBM Quantum 156-qubit Heron r2 quantum processors. Numbered dots are qubits and lines indicate couplings between qubits. 
    \textbf{(b)} Distance $d=3$ memory experiment patch for the heavy-hex code. $X$-type ($Z$-type) checks are shown in blue (red). Data qubits are yellow, measure qubits for $X$ check ($Z$ check) measurements are blue (red). Other qubits are in grey and those that are unused for the memory experiment are crossed (remnant flag qubits are grey un-crossed; see Methods). Data qubits are coupled to blue measure qubits for $X$ check readout via the red or grey qubits (see Methods). Logical $\bar{X}$ and $\bar{Z}$ operators for the patch are defined along the blue and red strings shown, respectively. Stabilizers for this code are formed by multiplying two $X$ checks (opposite $Z$ checks) together in a row (square).
    \textbf{(c)} A stability experiment patch for the heavy-hex code. Each qubit intersects with two blue checks, so that the product of all blue checks is even parity. The patch shown has four $X$ stabilizers: two 2-body checks (that are also stabilizers) at the top and bottom, and two 5-body operators that are each the product of two $X$ checks. The $X$ stabilizer outcomes are all initially random. We perform error correction/decoding using the standard stim \cite{Gidney2021a} and PyMatching \cite{higgott2023} libraries, making detectors in the standard way.
    \textbf{(d)} Topological spacetime diagram of lattice surgery. Two logical patches with space-like code distance $d$ incur an entangling lattice surgery operation for $t$ rounds. The `memory' part of the experiment is highlighted in light green and the `stability' part in light orange. The grey membrane represents the logical operator measured by the lattice surgery. In the memory (stability) part, a logical error occurs if a string of errors runs between opposite spatial (temporal) boundaries, intersecting the grey membrane, as shown by the dashed horizontal (vertical) line.  
    \textbf{(e)}    Diagram of the stability experiment over 6 rounds with qubits initially reset in the \(Z\) basis.
    Every round, the same \(X\)-type detector,  composed of the two \(X\)-type checks in the row, is afflicted by a measurement error (green cross).
    This measurement error is not detected and flips the logical (product of blue checks).
     \textbf{(f)} The original heavy-hex syndrome extraction circuit where $X$ and $Z$ stabiliser information is collected in separate time-steps and where resets are performed. The total time to implement this circuit is $11.1\mu s$. The top line represents a measurement qubit used firstly in the $X$ stabiliser measurement and then reused to collect flag information when the $Z$ stabiliser is measured. The bottom qubit is used to measure the $Z$ stabiliser only. Data qubits used are represented by the four central lines.  The operators $U_1$ and $U_2$ represent two different syndrome extraction operations. 
     \textbf{(g)} The improved heavy-hex syndrome extraction circuit where $X$ and $Z$ stabilisers can be measured simultaneously, decreasing the total circuit runtime to $3.2 \mu s$. The operator $U_3$ has a longer runtime that $U_1$ and $U_2$ in (f). 
     \textbf{(h)} The circuit in (g) with resets has a runtime of $5.4 \mu s$.
    \label{fig:devicelayout}}

\end{figure*}

\subsection{A heavy-hex code on a quantum chip}

We consider the heavy-hex code~\cite{chamberland2020}, realised on the IBM Quantum 156-qubit Heron class of quantum processor \textit{Marrakesh}. This device has a heavy-hex layout, displayed in \cref{fig:devicelayout}(a). It supports mid-circuit measurements and high fidelity ($> 99\%$) two-qubit gates. This generation of device, with its size, fidelity, and capabilities opens up the possibility for in-depth analysis and characterization of logical operations on error-corrected qubits.

The heavy-hex code is placed on the quantum processor as shown in \cref{fig:devicelayout}(a) to maximise performance. 
Detailed error modelling of the full device, including mid-circuit measurements using simultaneous randomised benchmarking informed our choice of code placement (see \cref{sec:characterisation}). 
Marrakesh supports a sufficiently large patch of high-quality qubits to support a $d=3$ logical qubit.
We run two experiments---memory and stability---that together help determine the capabilities of the device for executing quantum logic via lattice surgery.

\begin{figure}
\begin{tikzpicture}
\node[inner sep=0pt] at (-4.5,0)         {\includegraphics[width=1\linewidth]{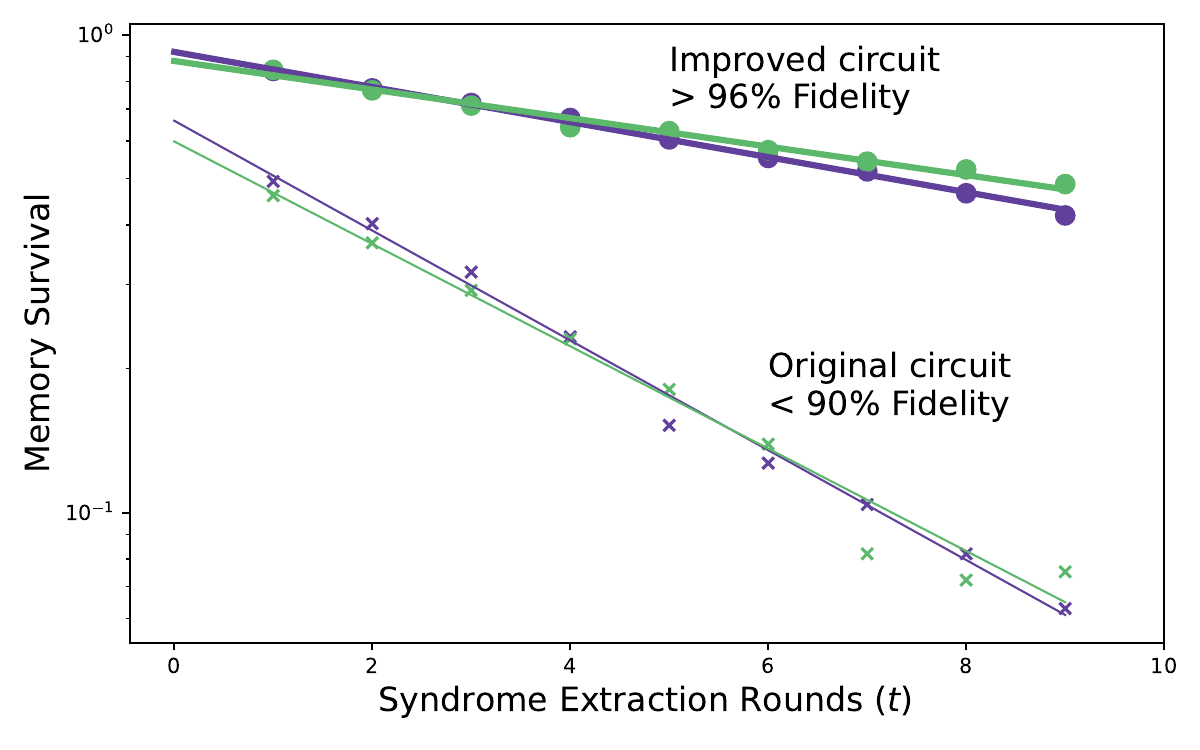}};
\node [shape=rectangle,align=center,fill=white] at (-6.2,-1.1) {{\footnotesize
        \begin{tabular}{lcc} 
            \multicolumn{3}{c}{\textbf{Improved Circuit}}\\
            \multicolumn{2}{c}{Basis}&Fidelity\\
            \begin{tikzpicture}
                \filldraw[purple_plot] (0,0) circle (2pt);
            \end{tikzpicture}&$|0\rangle$& 96\%\\
            \begin{tikzpicture}
                \filldraw[green_plot] (0,0) circle (2pt);
            \end{tikzpicture}&$|+\rangle$ & 97\%\\
        \end{tabular}}
    };
\end{tikzpicture}
    \caption{ \textbf{Memory experiment} using the $d=3$ heavy-hex code on the IBM Quantum processor \textit{Marrakesh}. Here we compare two syndrome extraction circuits. The original circuits are those used to implement the heavy-hex code as detailed in Ref.~\cite{sundaresan2022}, requiring $X$ and $Z$ checks to be measured and reset in separate time-steps. The improved circuits are introduced in \cref{sec:newcode}. For our new circuits, both $X$ and $Z$ checks are measured in the same time-step without post-measurement resets.
    Purple (green)  data indicates the system is initialised in the $|0\rangle$ ($|+\rangle$) state. There was negligible difference in the error rates if the logical qubits were initialised in the $|1\rangle$ ($|-\rangle$) state (data not shown).
    Here we fit the data (the logical success probability after $t$ rounds) to $Ap^t+0.5$, where $A$ is a SPAM parameter and $p$ is the decay factor. The logical fidelity is then $(1+p)/2$. We plot memory survival, which is a rescaled y-axis,  so that we are plotting $2p-1$. This re-scales the survival probability to be from 1 to 0, giving straight lines with the semi-log plot.}
    \label{fig:mresults}
\end{figure}

\subsection{Memory experiment}

Performing fault-tolerant logic requires keeping logical qubits uncorrupted over many error-correction cycles. A memory experiment quantifies how well logical quantum information can be stored by a code over time, measured in the number of syndrome extraction rounds. Memory experiments exploring logical failure rate as a function of the number of rounds have been detailed previously in superconducting devices for a number of different codes, including the heavy-hex code~\cite{sundaresan2022}, the surface code on heavy-hex lattices \cite{bence2024}, and the surface code on 4-valent lattices~\cite{google2022,google24}. 

We run two types of memory experiment. The first is based on the standard implementation of the heavy-hex code, wherein $Z$ and $X$ checks are measured in two separate rounds, with measure qubits measured and then reset in each round~\cite{chamberland2020}. This two-round approach results in poor memory performance owing to the time taken to perform resets and measurements, during which idling errors affect all data qubits (see \cref{sec:newcode}).

We redesign the syndrome extraction circuits to allow for all checks ($Z$ and $X$ type) to be measured in the same round (see \cref{sec:new_circuits}) and we remove resets from the circuits (see \cref{sec:reset_discussion}). We present the time savings of these new circuits schematically in \Cref{fig:devicelayout}(f).

\Cref{fig:mresults} details the results of our memory experiments on \textit{Marrakesh}, for both original and improved syndrome extraction circuits.
We use a standard minimum weight perfect matching decoder (PyMatching)~\cite{higgott2023}, populated with averaged device calibration data, and we do not post-select any data.  Using the improved circuits allows us to increase the logical fidelity per syndrome extraction round, from less than 90\% using the original syndrome extraction circuit, to 96\% and better using the improved circuit.
Based on our modelling, the main source of logical infidelity is the relaxation of qubits while mid-circuit measurements are performed (see \cref{sec:midSingleProtocol}). 

\begin{figure}
\begin{tikzpicture}

\node[inner sep=0pt] at (0,0)         {\includegraphics[width=1\linewidth]{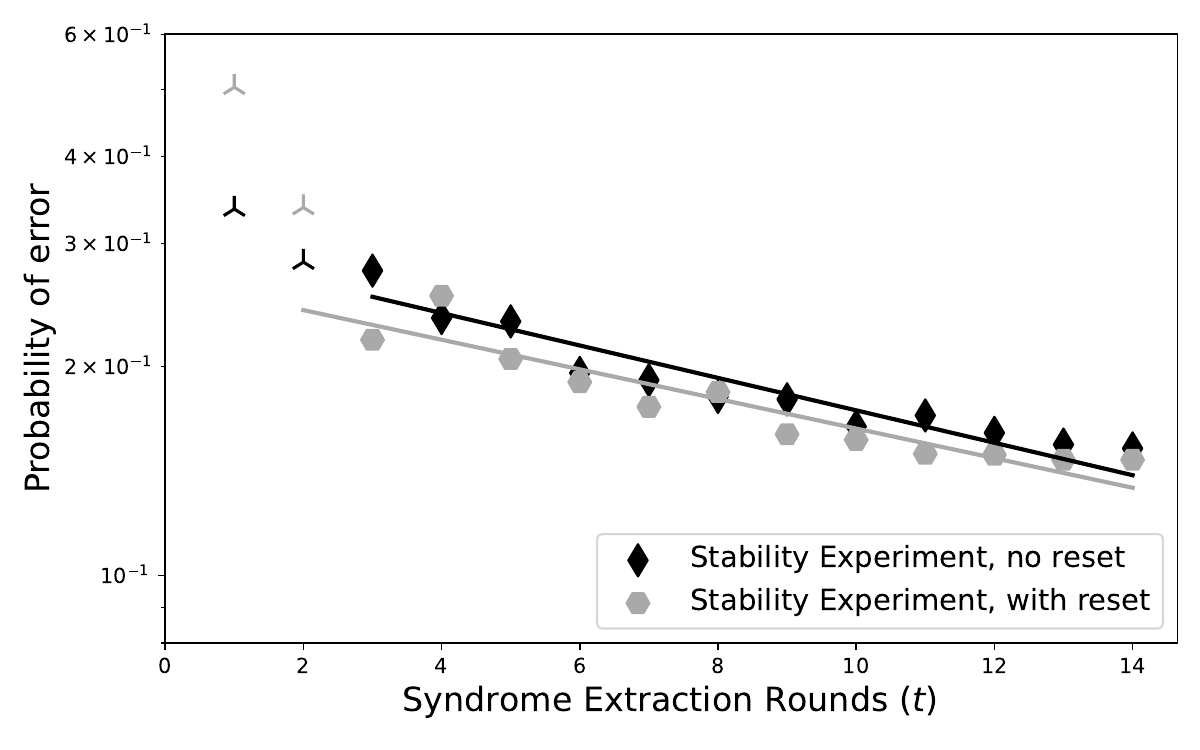}};

\end{tikzpicture}
    \caption{ 
    \textbf{Stability experiment} for the heavy hex code patch on the IBM Quantum processor \textit{Marrakesh}. Here we compare reset and no reset syndrome extraction circuits. Increasing the number of rounds leads to a decrease in the logical error probability, indicating we are below threshold for the class of errors detected by this experiment. 
    The first two rounds of syndrome extraction do not have full stabilizer information and, as indicated by the different symbols, were excluded from the fit. The data were fit to a simple exponential decay curve. 
    \label{fig:sresults}}
\end{figure}

\subsection{Stability experiment}


Given that we can store logical qubits, our device requires additional functionality to perform logic gates. 
We focus on an approach to performing fault-tolerant operations via lattice surgery, which is driven by performing additional stabilizer measurements.  
During lattice surgery, the value of a product of many stabilizer measurements must be read out correctly to successfully complete a logical operation.
If errors cause an incorrect stabilizer measurement outcome, the product will also be incorrect, and this will lead to gate failure. 
This can be mitigated by repeating stabilizer measurements for a number of rounds $t$, as in \cref{fig:devicelayout}(d). 
A stability experiment~\cite{Gidney2022dual} tests a device's ability to successfully measure products of stabilizers, and so serves as a proxy for logic gate performance.

In a stability experiment the code is re-designed with an over-complete set of stabilizer checks such that the product of outcomes of the over-complete checks are constrained to give a fixed $+1$ outcome. This constraint will only be violated due to undetected measurement errors. In \cref{fig:devicelayout}(c) we show how to modify the heavy-hex code to introduce one such constraint.

We detect measurement errors by comparing repetitions of stabilizer measurements. A detection event is defined where two consecutive readings of the same stabilizer do not agree. These detection events are fed to a decoding algorithm to attempt to recover the correct value for logical observables or, in the case of our stability experiment, the stabilizer constraint.

A measurement error produces two concurrent detection events. An undetectable failure of the stability experiment will occur if, say, one specific stabilizer fails at every repeated round of syndrome extraction, such that no detection events are identified. We decrease the likelihood of a logical failure by repeating syndrome extraction over more rounds. An example of such a logical failure in a stability experiment for the heavy-hex code is shown in \cref{fig:devicelayout}(e). 

We compare the performance of two different syndrome extraction circuits using the stability experiment, where in one variation we remove the reset from the circuit.
As discussed in Ref.~\cite{geher2024} and in \cref{sec:reset_discussion}, this requires forming detectors from stabilizer outcomes not in consecutive measurement rounds, but in rounds separated by two. 
The net effect of this change is to reduce the time-like distance of the code, but it also reduces the number of potential errors (since there are no reset errors). We discuss the trade-offs associated with including resets in \cref{sec:reset_discussion}.

\Cref{fig:sresults} shows the results of the stability experiment implemented using the improved circuits (see \cref{Sec:Methods} for implementation details), with and without resets. 
The stability experiment failure rate as a function of the number of measurement rounds can be modelled to leading order by a simple exponential $P_{\mathrm{fail}} = B(d) \Gamma^t$ where $B(d)$ represents an unknown pre-factor polynomial in the code distance $d$, and $t$ is the number of syndrome rounds. We fit an exponential to these data in \cref{fig:sresults}, omitting the initial rounds (the small data points in the figure) from the fit.  We observe only a negligible difference between our two syndrome extraction circuits. A likely explanation is that resets in the \textit{Marrakesh} device are performed via measurements followed by a conditional $X$-gate. The resets, therefore, have similar error rates to the measurements, and although the circuit with resets has twice the time-like distance, we have also substantially increased the chance of errors -- more or less cancelling out the benefit. We note that after more than about 15 rounds the experimental data points in \Cref{fig:sresults} appear to flatten. This does not occur in our simulation and is likely indicative of additional noise processes. 

\subsection{Impact of noise parameters on stability and memory}
\label{sec:memoryandstabilitytradeoff}

\begin{figure}
    \centering
    \includegraphics[width=0.5\textwidth]{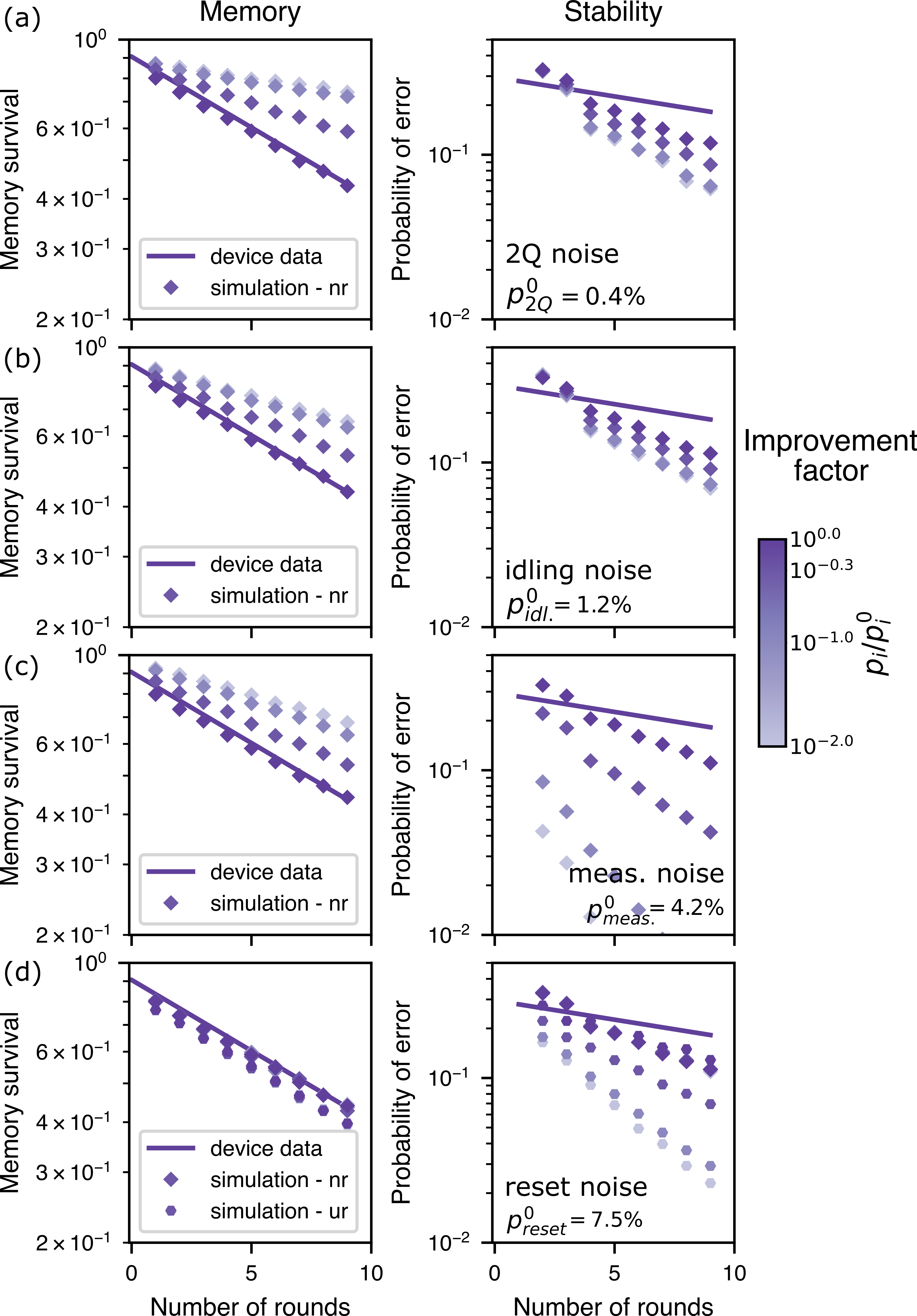}
    \caption{\textbf{Simulated effect of circuit noise parameters} on memory (left) and stability (right) experiments. For the memory experiment, we calculate the memory survival as $1-2\bar{p}_{\rm fail}$ with $\bar{p}_{\rm fail}$ the probability of error. With this definition, the memory survival is expected to take the form $Ap^t$. Experimentally measured data are fit to this expected scaling behaviour as in \cref{fig:mresults,fig:sresults} (thick line).   Logical failure rates were calculated in Stim~\cite{Gidney2021a}, with no reset (`nr' - diamond) and unconditional reset (`ur' - circle). Each noise parameter is varied independently, improving noise from a value $p_i^0$ determined by a fit to the experimental data (dark) to up to two order magnitude improvement (light) in steps (1/2, 1/10, 1/100) as shown on the common color scale. \textbf{(a)} Effect of two qubit error rates. This benefits both memory and stability circuits up to $p_{\mathrm{2Q}} \sim p_{\mathrm{2Q}}^0/10$. \textbf{(b)} Effect of idling noise on data and measurement qubits (we fixed $p_{\mathrm{idl.}}=p_{\rm q.meas}$). This improves both experiments similarly as with two qubit error rates, also up to $p_{\mathrm{idl.}}\sim p_{\mathrm{idl.}}^0/10$. \textbf{(c)} Effect of measurement noise. Reduction of measurement noise strongly improves the stability experiment, which is expected since measurement errors are the underlying mechanism causing logical errors in stability. It also decreases the logical failure rate of the memory, although less dramatically. This parameter has the biggest effect on the logical failure probability, suggesting it is the dominant source of noise. \textbf{(d)} Effect of reset noise. We add reset for both circuits and observe a reduction in logical errors only for the stability `ur' circuit. As expected the memory experiment was fairly insensitive to the quality of the reset operations \cite{geher2024}. Even with the stability circuit, improving reset does not allow one to achieve the same level of error reduction as one gains by improving measurement.}
    \label{fig:noisemodel_effect}
\end{figure}

Here we use simulations to examine the impact of varying the noise parameters of various constituents of the quantum processor on the performance of stability and memory circuits. We fit the device noise to a circuit-level noise model involving 1- and 2-qubit depolarising noise, measurement and reset errors and idling noise (see Section~\ref{sec:Simulation_details}). The best-fit parameters for the device are included in \cref{fig:noisemodel_effect}. We also vary these parameters individually to examine the effect of each noise parameter on stability and memory performance and present these results in \cref{fig:noisemodel_effect}.

As we see in \cref{fig:noisemodel_effect}(c), improving measurement noise is predicted from our simulations to have the largest effect on stability performance, and is also predicted to improve memory performance considerably. This suggests that our results are currently limited by measurement noise. 


\section{Discussion}

Successful quantum computation will require hardware to maintain quantum information in both space and time, keeping logical qubits stored in memory alive while logical gate operations occur. 
For lattice surgery operations, we require the ability to maintain logical qubits in memory over several rounds of syndrome measurement, during which time we must successfully read out stabilizer outcomes that drive logic gates. 
These dual challenges are characterised by memory and stability experiments, respectively. In this work, we have presented the results of both experiments on a superconducting qubit array, for the first time characterising these two aspects of fault-tolerant quantum computing performance on the same device. 
We optimise the results of these experiments by re-designing syndrome measurement circuits for the heavy hex code, removing the need for slow and noisy resets, and by positioning the code patches in an optimal area, as identified by extensive device benchmarking.

Ultimately, to maximise the performance of a logical entangling operation, one must maximise the combined probability of success for both the quantum memory and the stabilizer readout (the stability part of Fig.~\ref{fig:devicelayout}(d)). 
This will involve tuning the distance $d$, quantifying the number of errors that the code can correct, and $t$, the number of rounds over which one performs lattice surgery measurements.
The performance of a stability experiment will worsen as the size of the patch is increased, just as the memory fidelity worsens with increasing number of rounds. 
However, stability (resp.\ memory) performance is expected to improve exponentially with $t$ (resp.\ $d$) and only decrease as some polynomial in $d$ (resp. $t$). Hence, one can hope to improve the logical gate fidelity by increasing both $t$ and $d$.
For each $d$, there will be some optimal value of $t$ at which the trade-off between memory and stability is balanced, and the overall probability of success for the logical operation is maximised, and this optimum will be dependent on device noise parameters. 
A focus of the next generation of fault-tolerance experiments should be determining these optimal logical gate parameters and, hence, characterising the maximum success probabilities of logical operations.

It is worth noting that, for the heavy-hex code, we do not expect the memory results to improve arbitrarily for larger $d$, since the code does not possess a threshold for these experiments. 
Different codes, such as Floquet codes or surface codes adapted to the heavy-hex lattice, could provide suitable alternatives while remaining compatible with the layout of quantum processors such as these. 
Including resets in memory and stability circuits is likely to degrade memory performance with no improvement in stability performance in future experiments unless the current resets are replaced with resets that cause significantly less noise than the mid-circuit measurements~\cite{geher2024}.

In this work, we have identified measurement noise as a key limitation on the performance of lattice surgery operations in present-day devices, and this needs to be improved to reach the logic gate fidelities at which we will need to be operating in a large-scale quantum processor. 
We have shown that, while the device tested is showing improving stability performance with the number of rounds, improved mid-circuit measurements would dramatically enhance stability performance, and hence, the success probability of logical operations.

\section{Methods}
\label{Sec:Methods}

\subsection{Implementation of the Heavy-Hex Code}\label{sec:newcode}

\subsubsection{Introduction to Heavy-Hex Code}

The heavy-hex code is a subsystem code defined by a ``gauge group", which is generated by weight-2 $Z$ checks and weight-4 (in the bulk of the code) $X$ checks. 
These are shown in red and blue, respectively, in \cref{fig:devicelayout}(b). 
As usual in a subsystem code, the checks we measure directly are not the stabilizers of the code. 
Rather the stabilizers are those gauge group elements that commute with all measured checks. 
The product of two vertically opposite $Z$ checks forms such a stabilizer, as does the product of all $X$ checks along a row. 
In \cref{fig:devicelayout}(b), there are just two $Z$ stabilizers and four $X$ stabilizers (two in the bulk and two on the top and bottom boundaries). 
The values of these stabilizers are inferred from the measurements of their comprising checks. 
With the boundary conditions shown in \cref{fig:devicelayout}(b), this code stores a single logical qubit, with ``bare" $X$ and $Z$ logical operators shown, which commute with all the checks. 

In the original formulation of the heavy-hex code, the checks are measured in two rounds~\cite{chamberland2020}. 
In one round, the $X$ checks are measured with a syndrome measurement circuit that couples the data qubits to the green ancilla qubits via the purple qubits, which act as ``flag" qubits. 
The green and purple qubits are measured at this point, with the flag qubits providing extra information on the likelihood of hook errors having occurred, which is fed to the decoder~\cite{chamberland2020}. 
In the next round, the $Z$ checks are measured using the purple qubits as ancillas. 
We redesign these circuits so that they can be executed in a single round (without flags in the bulk of the patch), thereby reducing the overall time required for syndrome extraction. We introduce this circuit below.

\subsubsection{Need to redesign the code}
A distance-$3$ heavy-hex code has been previously implemented on an IBM Quantum processor in Ref.~\cite{sundaresan2022}. 
This implementation used the 27-qubit quantum processor \textit{Peekskill}, which was a fixed qubit, fixed coupler device (like the current Eagle-class processors) and is no longer available. 
In Ref.~\cite{sundaresan2022} the authors achieved logical errors for a logical qubit in the $\ket{0}/\ket{1}$ basis of under 6\% and in the $\ket{+}/\ket{-}$ basis of under 12\%. 
Where they post-selected for leakage error the maximum logical error fell to under 8.6\%. 

IBM's current flagship quantum processors are of the Heron class, which are fixed qubit, tunable coupler devices. 
On the r2 devices, two qubit error rates have improved from a median error rate of approximately 2\% on current Eagle-class devices to a median error rate of approximately 0.35\% on Heron-class devices, representing nearly an order of magnitude improvement. 
These error rate improvements appear to come at the cost of slightly shorter $T_1$/$T_2$ times, which are on the order of 213/120~\textmu s on the best Heron device. 
At the same time, the duration of a measurement has increased, from a combined measurement and reset of 768 ns to between 3000 and 4000 ns, depending on which device is used. 
For mid-circuit measurements, this is a substantial fraction of the $T_1$/$T_2$ time, and leads to worse logical qubit performance than that reported on \textit{Peekskill}, despite the improved two-qubit gate operations. 
The timing problems are further exacerbated by the fact that the original heavy-hex code extracted $Z$-type stabilizers and $X$-type stabilizers in separate syndrome extraction circuits, executed in series (one after the other), requiring an idle time on non-measured qubits of over 8 \textmu s per round of syndrome extraction.  
We provide details of how we measured the noise impact of performing measurements mid-circuit in \cref{sec:midSingleProtocol}.

\subsubsection{Improved syndrome extraction circuit}\label{sec:new_circuits}
\label{sec:improved}

\begin{figure*}
    \begin{tikzpicture}
    \node[inner sep=0pt] at (-9,4)         {\includegraphics[scale=0.5]{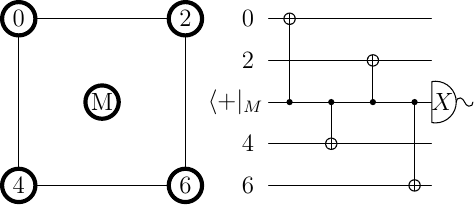}};
     \node[inner sep=0pt] at (-4,4)         {\includegraphics[scale=0.5]{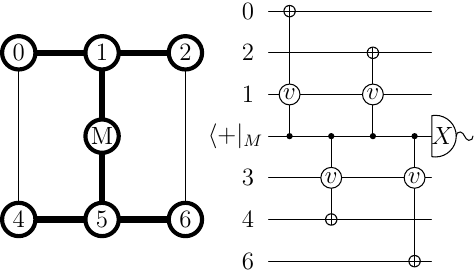}};
     \node[inner sep=0pt] at (-9,1.5)         {\includegraphics[scale=0.65]{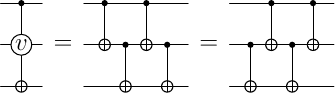}};
      \node[inner sep=0pt] at (2.7,1.5)         { \includegraphics[width=0.35\textwidth]{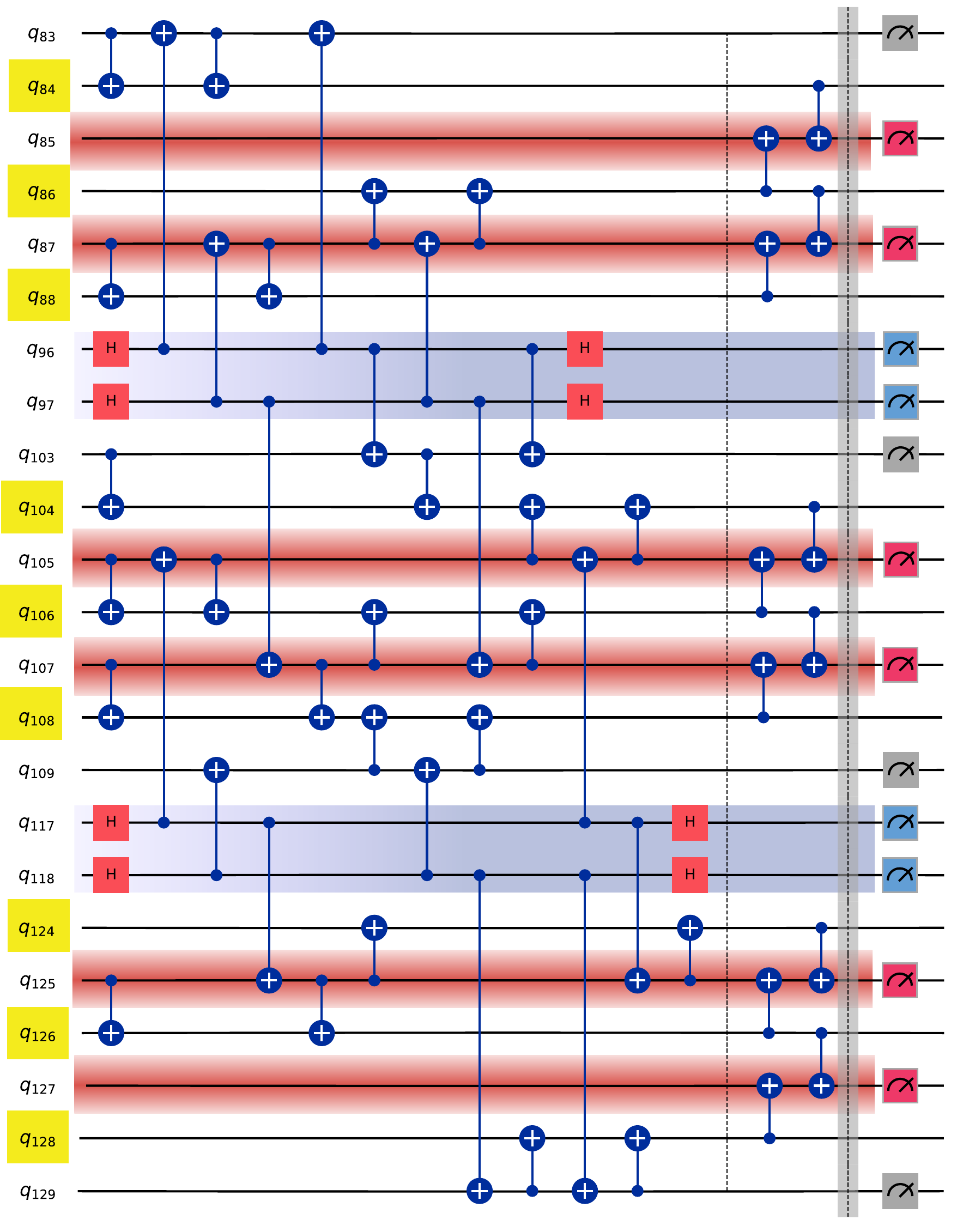}};

    \node[inner sep=0pt] at (-3.5,1.2)         {\includegraphics[scale=0.4]{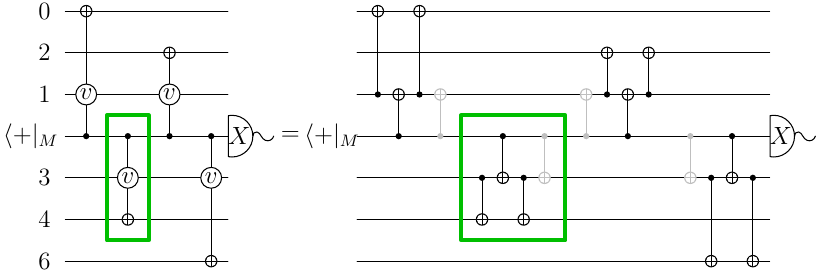}};
    \node[inner sep=0pt] at (-6.5,-1.5)       {\includegraphics[scale=0.6]{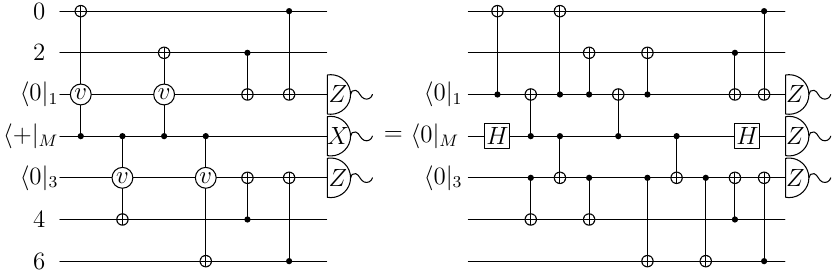}};
    \node[inner sep=0pt] at (-11,5.3)         {(a)};
    \node[inner sep=0pt] at (-6.8,5.3)         {(b)};
    \node[inner sep=0pt] at (-11,2.6)         {(c)};
    \node[inner sep=0pt] at (-6.5,2.6)         {(d)};
    \node[inner sep=0pt] at (-11,-0.1)         {(e)};
    \node[inner sep=0pt] at (-1,5.3)         {(f)};
    \end{tikzpicture}
    \caption{\textbf{Designing the new syndrome extraction circuits.} 
    \textbf{(a)} A circuit to measure a weight-four stabilizer. The ordering is chosen to minimise circuit depth and to mitigate the effect of hook errors. Although the circuit can introduce weight-two errors on the data qubits, they are aligned such that their effect is relatively benign.
    \textbf{(b)}  A weight-four check on the heavy-hex architecture. We measure a weight-four stabilizer on the data qubits (qubits with even index). However, the measurement qubit $M$ is separated from the data qubits via an additional qubit designed for use as a flag qubit. We attempt to recover the readout strategy shown in (a)  using a next-nearest-neighbour controlled-not gate. Where the control and target qubit share a neighbouring `via' qubit. We mark the qubit with a $v$ in the circuit diagram to the right. 
    \textbf{(c)} Circuits to realise next-nearest-neighbour CNOT gates in terms of four nearest neighbour CNOT gates. 
    \textbf{(d)} The weight-four parity check circuit expanded in terms of nearest-neighbour gates. We show the expansion of one next-nearest-neighbour gate explicitly in the green box. With an appropriate choice of expansion we find certain physical gates can be cancelled using standard relations between gates. Gates that are cancelled are shown in grey. 
    \textbf{(e)} An expanded circuit to measure both a weight-four Pauli-Z check and two weight-two Pauli-X checks on a single plaquette. The circuit has depth twelve including qubit initialisation and readout and can be applied in parallel on all (even) plaquettes, thereby enabling the extraction of the full syndrome within a single round of measurements. 
    \textbf{(f)} Circuit diagram for the new logical heavy-hex code (see \cref{sec:newcode} for details). Data qubits are marked in gold and the red/blue measurement qubits are highlighted. The initial part of the circuit (up to the dotted line) represent preparation of the blue stabilizers (here X-stabilizers) and after the dotted line the red stabilizers (here Z-stabilizers). All measurement qubits are measured in the same time step. The qubits measured each round that are not measurement qubits are remnant flag qubits, that can be used to provide further information to the decoder.
}
    \label{fig:newcodeLayout}
\end{figure*}
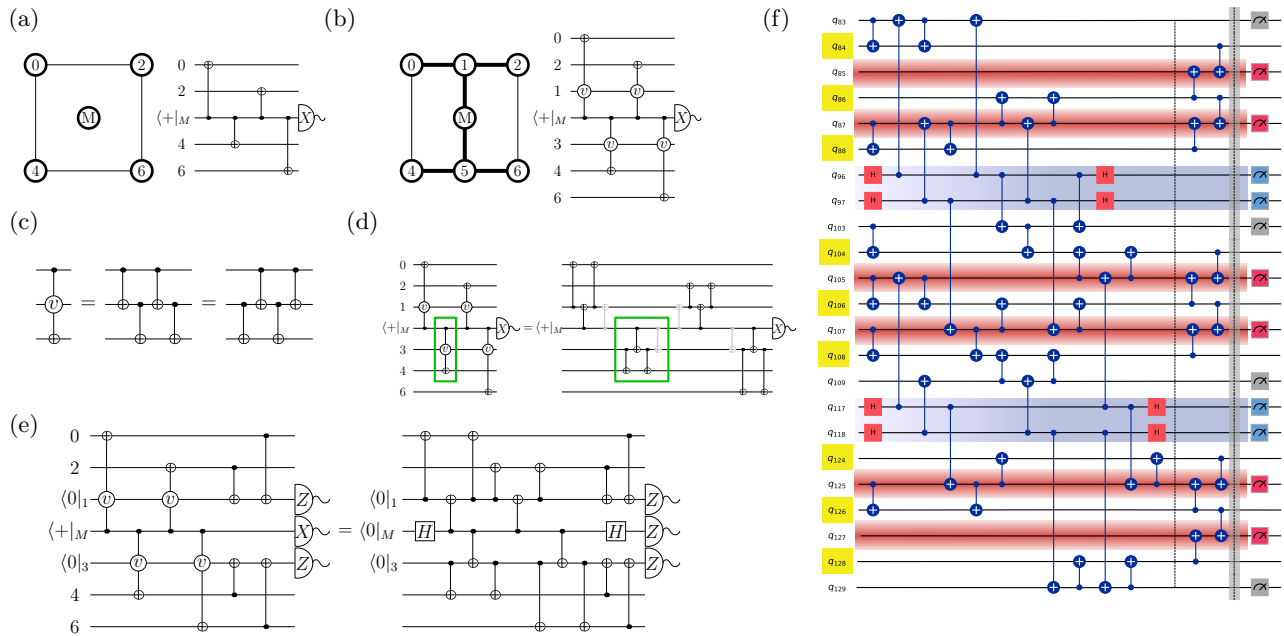

Here we introduce a variant of the syndrome extraction circuit for the heavy-hex code, that simultaneously prepares both the $X$ and $Z$ stabilizer ancillas while avoiding the hook errors that would normally require the use of flag qubits. The full circuit used is shown in  \cref{fig:newcodeLayout}(f).

Let us give some intuition for the theory behind this circuit. Both the heavy-hex code and the surface code make weight-four stabilizer parity measurements in their syndrome extraction circuit. However, the heavy-hex lattice was designed to read out the weight-four checks of the heavy-hex code using flag qubits to mitigate the effect of hook errors. These flag qubits are also used in later rounds of measurements to measure other stabilizer operators. As these qubits are used for two purposes, the complete syndrome readout circuit requires two rounds of measurements. Using intuition from the `standard' surface code readout circuit, we can avoid the need for flag qubits on the heavy-hex lattice, such that we can measure all checks in the same round of measurements.

The surface code realised on the square lattice can avoid the use of flag qubits with an appropriate choice of syndrome readout circuit, see \cref{fig:newcodeLayout}(a). We find that adopting this circuit on the heavy-hex lattice using next-nearest-neighbour CNOT gates gives us a syndrome extraction circuit on the distance-three code that can identify any single error, and measure all types of checks simultaneously. See the circuit in \cref{fig:newcodeLayout}(b). Here we mimic the surface code stabilizer check using next-nearest-neighbour CNOT gates. \Cref{fig:newcodeLayout}(c) shows the circuit identities used \cite{bence2024}, which we can adopt here. 

In \cref{fig:newcodeLayout}(d) we show the circuit that measures all the checks for the syndrome extraction circuit including a weight-four Pauli-Z check and two weight-two Pauli-X checks. 
This circuit is expanded in \cref{fig:newcodeLayout}(e). 
The total circuit uses ten layers of one-and-two qubit unitary gates in between qubit reset and readout. 
This circuit is parallelizable over all plaquettes due to the fact that qubits on opposite corners of the plaquette are never used simultaneously. This means adjacent plaquettes that share a corner do not compete to use the same qubit at the same time. For instance, qubit 0 is never addressed at the same time as qubit 6 and likewise qubit 2 is never addressed at the same time as qubit 4.

\subsubsection{Eliminating the need for reset}\label{sec:reset_discussion}

An additional improvement we can make to the syndrome extraction circuit of the heavy-hex surface code is to eliminate resets of the measurement qubits. Here we follow the ideas outlined in Ref.~\cite{geher2024}. Ordinarily, syndrome extraction circuits measure and then reset qubits.
Resetting qubits after measurement serves to decorrelate errors in the measurement outcome and quantum state of the qubits. This otherwise results in time-like correlated errors that serve to reduce the distance of stability experiments~\cite{geher2024}.

The \textit{Marrakesh} device employs conditional reset, which consists of a measurement and a Pauli $X$ gate conditioned on the measurement outcome. 
As this Pauli $X$ gate can be tracked in software, measuring and then resetting qubits is equivalent to simply measuring twice. 
The device is in the regime in which the measurement dominates the time needed for a round of syndrome extraction and hence, we seek to remove unnecessary measurements from the circuit.
We can indeed modify detectors to accommodate syndrome extraction circuits without reset or, equivalently, a second measurement~\cite{geher2024}.
The only modifications required are to the detectors. During syndrome extraction, detectors must compare measurement outcomes with two rounds prior, rather than the prior round. 
Initial and final detectors differ from this pattern (final detectors need to take into account the state of the previous detector). PyMatching can decode the resulting experiments with no appreciable loss in decoding power. 
This modification halves the time-like code distance for stability experiments, since flipping every second measurement outcome for a single check can result in an undetected logical failure. 
But when we compare this to a circuit using reset (implemented by measurement), it also halves the number of measurements and the noise associated with these measurements. 

\subsection{Using simulation to examine memory/stability tradeoff}\label{sec:Simulation_details}

We simulate the stability and memory circuits using stim \cite{Gidney2021a} to reconstruct a potential noise model for the observed data from the device, and to predict the effect that improving various aspect of the noise would have on each of these experimental tests.

The noise model we explore is defined as follows. Before each qubit operation on the data qubit, we add a depolarizing channel with parameter $p_{1Q}$ for single qubit operation and $p_{2Q}$ for two qubit operation. Inspired by Ref.~\cite{geher2024}, we separate measurement noise into a quantum and a classical part. 
The quantum part corresponds to applying a Pauli-X  to the measurement qubit before measurement with probability $p_{\mathrm{q.meas}}$. The classical part accounts for the case where the measurement device outputs a wrong value independently of the state of the measurement qubit. For example, a classical error would lead to a measurement reading 0 even though the measurement qubit is in state $\ket{1}$. 
We parameterize this classical measurement error with $p_{\mathrm{c.meas}}$. 
Finally, we consider idling noise on data qubits when measurement and reset operations take place. An initial estimate of the idling noise can be made following the methodology of Refs.~\cite{rost2020simulationthermalrelaxationspin,geher2024} which parametrises idling noise with relaxation time $T_1$, dephasing time $T_2$ and measurement time $t$. On \textit{Marrakesh}, the quantum device used for our experiments, we find $T_1\simeq T_2$ and so consider the same level of noise for each Pauli channels $X$, $Y$, and $Z$. 
We therefore add a depolarizing channel with parameter $p_{\mathrm{idle}}$ on data qubits during measurement and reset operations. 

Finally, we define noise on reset by adding a Pauli-X on the qubits being reset after each reset operation, with probability $p_{\mathrm{reset}}$. In \cref{tab:noisemodel} we report the noise parameters obtained by numerically minimising the mismatch between simulated and experimental data for the memory and stability experiments where we have parametrized the model such that $p_{\mathrm{q.meas}} = p_{\mathrm{idle}}$. This means the quantum measurement noise is caused by the measurement qubit idling during the time of the measurement. These parameter choices are in agreement with device characterization data presented in Sec.~\ref{sec:characterisation}.

\begin{table}[h]
    \centering
    \begin{tabular}{l|l}
    \hline 
        $p_{1Q}$  & 0.02\%  \\
        $p_{2Q}$  & 0.41\% \\
        $p_{\mathrm{q.meas}}$ & 1.2\%  \\
        $p_{\mathrm{c.meas}}$ & 4.2\%  \\
        $p_{\mathrm{idle}}$ & 1.2\%  \\
        $p_{\mathrm{reset}}$ & 7.5\%  \\
    \end{tabular}
    \caption{Noise parameters obtained by fitting to experimental data. Parameters are found from numerical optimisation using the Nelder-Mead method. }
    \label{tab:noisemodel}
\end{table}

To examine the effect of these noise sources in both the memory and stability experiments,  we vary in simulation the physical error rate of the relevant noise sources; see Fig.~\ref{fig:noisemodel_effect}. 
Overall this analysis highlights the dominant role of measurement noise in the stability experiment. Classical measurement errors are particularly detrimental for stability experiments, while minimising idling noise is crucial for both the memory and the logical operation on logical qubits.  Potentially, the stability experiment could be improved by using resets, but the quality of resets must be high.

\subsection{Characterising the quantum processor}
\label{sec:characterisation}

To identify the best location for code placement on the device, and to inform detailed error models that allow us to simulate logical performance of the experiments presented here, we undertake extensive characterisation and benchmarking of the quantum processor.  A range of efficient characterisation tools have been developed recently that allow for the rapid acquisition of rich, detailed noise characterisation data for quantum devices consisting of hundreds of qubits.

\subsubsection{Simultaneous randomised benchmarking}\label{sec:initialProtocol}

\begin{figure*}
    \includegraphics[width=1\textwidth]{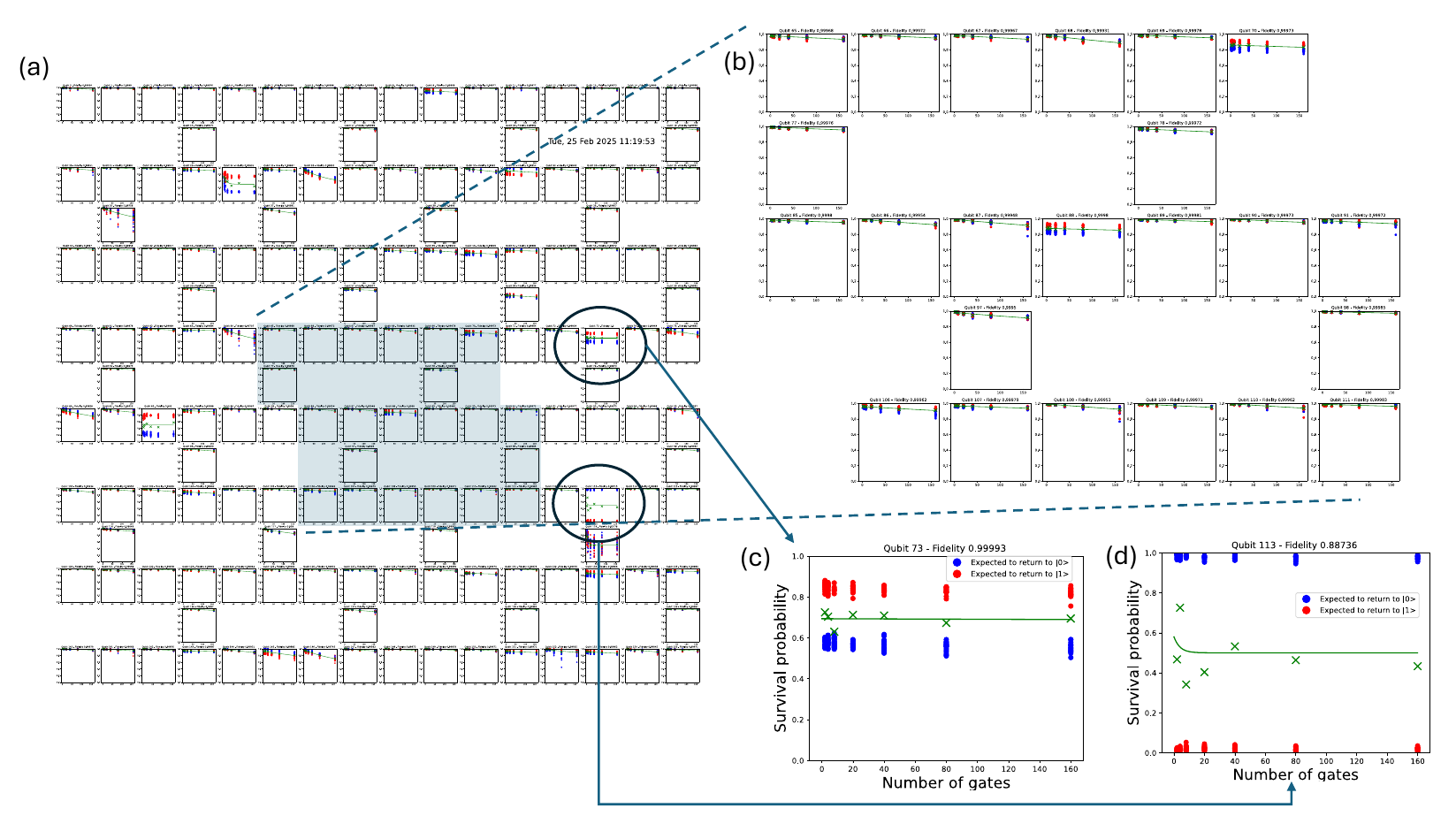}
    \caption{\textbf{Simultaneous randomised benchmarking.}  (a) Output from running our protocol for simultaneous RB.  Individual decay graphs for each qubit are arranged in the same location as in the device layout (the QPU time to gather this information is substantially less than a minute). We can use this type of visualization to quickly assess the state of the individual qubits, allowing us to identify required blocks of qubits. (b) An example of a block of qubits required to form a logical qubit for the heavy-hex code. (c) An example illustrating why it is important to plot the individual sequence runs as well as use a protocol that randomises the final state. Here the red dots represent the runs that are returned to the $|1\rangle$ state prior to final measurement, blue dots represent return to the $|0\rangle$ state. As can be seen from this example, there is a marked discrepancy between the accuracy of the measurements for different states. Also of import is that the fidelity reported by RB is still very high as RB is robust to SPAM. While the operation of the qubit is still high fidelity, in this case the SPAM error is so extreme that the qubit is probably not usable if it needs to be measured. (d) Another example illustrating extreme behavior. Here the qubit appears to read zero, whatever state the qubit is in. Clearly this behavior violates a number of RB assumptions, but a simple application of RB might yield high quality but misleading results.}
    \label{fig:single}
\end{figure*}

We can use randomized benchmarking (RB)~\cite{Dankert2009,Knill2008,Magesan2011} for an initial examination of the noise characteristics of the device. 
Here we use simultaneous single qubit Clifford twirls; see Fig.~\ref{fig:single}.
RB is made more effective if we use a variation of the protocol first suggested by Knill~\cite{Knill2008} and further analysed in Refs.~\cite{Harper2018} and~\cite{Andrews2019}. In this variation, rather than using strictly inverting sequences of gates, we further randomize whether an individual qubit ideally ends in $\ket{0}$ or $\ket{1}$.
By recasting the results as a survival probability (i.e., the probability of getting the expected result) we gain two immediate benefits: 
(1) by combining the results to obtain an average, we eliminate a nuisance parameter in the fitting procedure, increasing the accuracy of the results; and 
(2) by plotting the survival probability for each sequence using different colors to distinguish sequences where we expect to end in $\ket{0}$ and where we expect to end in $\ket{1}$ we can immediately detect measurement bias as well as potentially detecting measurement correlations across the device. 
The data from the experiment can also be analysed to detect correlations between the operation of the qubits \cite{Harper2019} (as distinct from purely measurement related correlations). 
We give an example of this below.

We use the RB protocol highlighted in~\cite{Harper2019}, but for ease of reference we summarize it here. 
Ref.~\cite{Harper2018} provides guidance as to sensible choices for the number of gates $m$ and the number of random sequences $k$ for a particular $m$. 
If we assume an $n$-qubit device, the exact randomized benchmarking variant used here is as follows:

\begin{enumerate}
    \item Choose a positive integer $m$. 
    This represents the number of twirling gates that will be performed in the sequence. 
    \item For each qubit, choose whether:
    \begin{itemize}
        \item to apply an initial $X$ gate (the sequence for that qubit will ideally return to $\ket{1}$) ; or
        \item not to apply an initial $X$ gate (the sequence will ideally return to $\ket{0}$).
    \end{itemize} Note which qubits had an $X$ gate applied ($x$).
    \item For each qubit, choose a sequence of $m$ random single-qubit Clifford gates and the Clifford gate that will invert that entire sequence (excluding any initial $X$ gate applied). 
    Apply these gate sequences in parallel.
    \item Sample the output of the circuit specified by 2 and 3 (referred to as $s$) by taking a number of shots. 
    For the protocol described here we don't need to retain the full set of bit patterns (although one could use them as described in~\cite{Harper2019}). 
    Rather, for each qubit ($q$) determine $\hat{p}(q,m,s,x)$, the observed probability that the qubit $q$ for that length ($m$) and sequence ($s$) was measured as returning to the expected state (determined by $x$). 
    For each qubit this is done by examining the relevant bit of the returned bit pattern for each shot and averaging the number of times the measurement was `successful' i.e.\ we measured a 0 if the qubit was supposed to return to $\ket{0}$ and we measured a 1 if the qubit was supposed to return to $\ket{1}$. 
    We refer to this as the \emph{survival probability}.
    
    \item Repeat steps 2-4 $k$ number of times to build up statistics.
    \item Repeat steps 1-5 for different choices of $m$. 
\end{enumerate}

The values of $m$ that are chosen will depend on the fidelity of the single qubit gates in the device when operated simultaneously. 
Assuming the fidelity of the device supports these numbers we want to begin at approximately $m=2$ or $3$ (\cite{Wallman2018,Merkel2021}). Ideally, the final $m$ should be chosen to return a `survival probability' of approx 60\% (and always substantially above the asymptote value of 50\%~\cite{Harper2018}), with several values in between. 
However, for systems with very high fidelity it might not be practical to select such a high final value of $m$. 
While this is not required, choosing a lower largest value will impact the `relative error' estimate of the fidelity. 

Under the assumptions of stationary Markovian noise with weak gate-dependence, the data marginalized to each qubit will yield an exponential decay of the survival probability of that qubit when averaged over each of the sequences. 

The data from such an experiment allows verification that the fidelity of the single qubit Clifford gates are in the expected region (e.g.\ similar to published calibration rates). 
For the sake of completeness the average gate fidelity for each qubit is obtained by averaging the results for each $k$ sequences $\hat{p}(q,m) = \frac{1}{k}\sum\hat{p}(q,m,s,x)$, where the sum is over each of the sequences for a particular length (measured in step 4 above). 
These averages are then separately fit for each qubit to the exponential decay curve $\hat{p}(q,m) = Ap_q^m + 0.5$ to determine an estimate of the decay constant $p$ for qubit $q$. 
(This is directly related to the average gate fidelity. Ref.~\cite{carignandugas2019} contains a useful table of conversions.)

Typically, with randomized benchmarking, it is only the average of the sequences $\hat{p}(q, m)$ that is used. 
There is a relationship between the coherence of the noise and the variance of the survival rate; see for example~\cite{Ball2016}. 
Certainly, purely depolarizing noise will have small variance and purely coherent noise will have wide variations in the survival rates of individual sequences. 
However, in both cases the average will, given the assumptions of randomized benchmarking, fit the single exponential decay curve (although, see Refs.~\cite{Fogarty2015,Ball2016,fong2017} for examples where non-exponential decay might be observed). 
Here we analyze the distribution of data from individual sequences, keeping track of which sequences were designed to return to $\ket{0}$ and those to $\ket{1}$. 
This provides us with a wealth of information, not only about the performance of the gates, but also about the bias of the measurement and, indeed, potentially correlations between measurements. 
We provide some examples below.

\subsubsection{Characterising mid-circuit measurements with simultaneous RB}\label{sec:midSingleProtocol}

\begin{figure*}
    \includegraphics[width=1\textwidth]{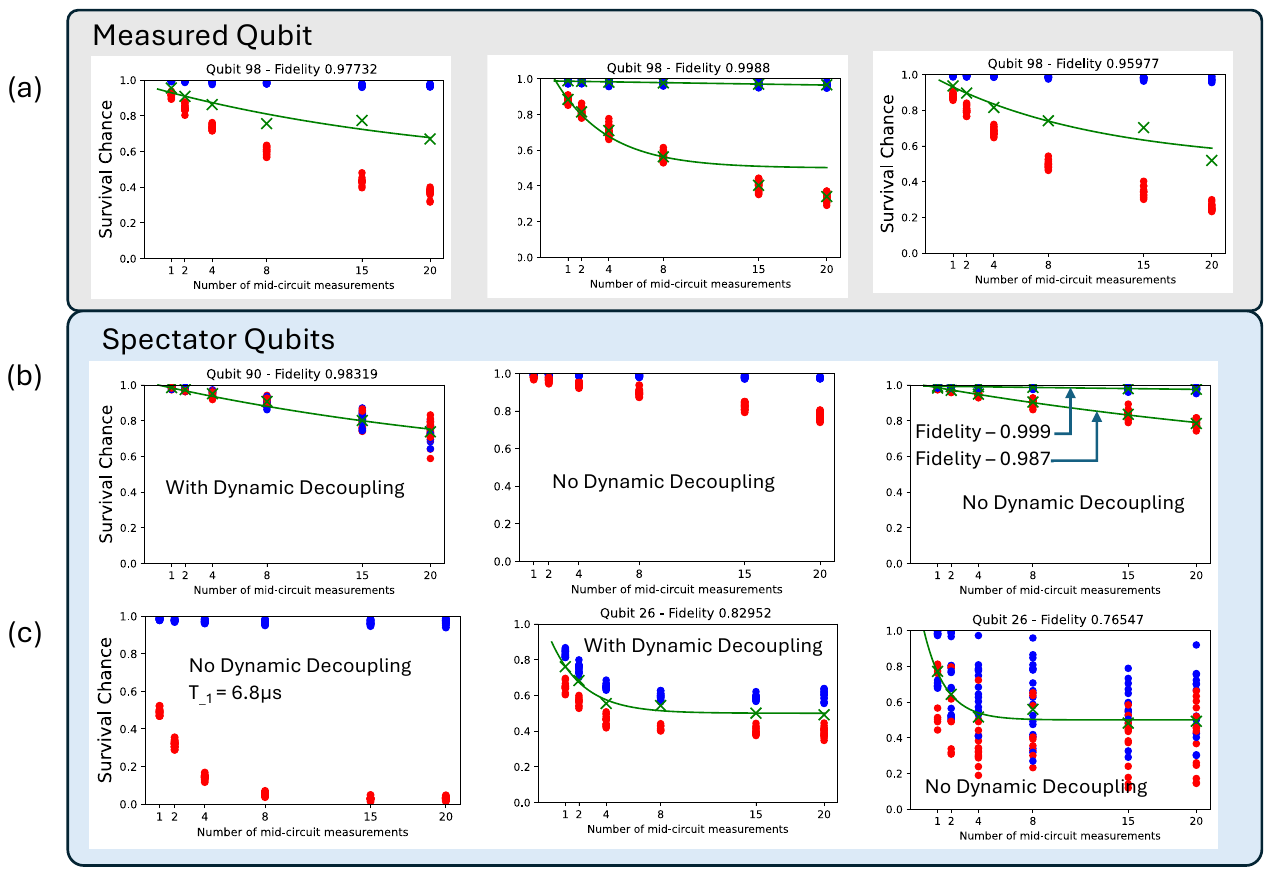}
   \caption{\textbf{The effect of mid-circuit measurements.}  \textbf{(a)} An RB experiment using a single Clifford twirling operation, interleaved with mid-circuit measurements. For this experiment we performed four single Clifford gates between each measurement. For each qubit being measured (a \textit{measurement qubit}) it was randomly determined if the qubit would be returned to the $|0\rangle$ state (blue dots) or the $|1\rangle$ state for the final measurement. Leftmost graph here we return to the qubit to the same state as the designated final state, prior to each mid-circuit measurement. Only the final measurement is shown. This is clearly not a typical RB decay curve. Of most import is that the space between the red dots (return to $|1\rangle$) and the blue dots (return to $|0\rangle$) increases with sequence length. Middle plot, this plot combines two different experiments. In the first, we change the protocol so that, regardless of the final measurement the qubit, is returned to the $|0\rangle$ state prior to a mid-circuit measurement  - blue dots. In the second the qubit is returned to the $|1\rangle$ state prior to the mid-circuit measurement - red dots. As can be seen the decay plots are mainly impacted by the state of the qubit during the mid-circuit measurement. Rightmost plot, here we repeat the first experiment, but instead of mid-circuit measurement we just introduce a delay of the same amount of time required for a mid-circuit measurement. This plot is strikingly similar to the left-most one, providing evidence that it is the delay that is causing a qubit in the $|1\rangle$ state to decay. \textbf{(b)} Left plot: Here we plot the decay curve of a spectator qubit, while the experiment discussed in (a) is being performed. The qubit is returned to the same state it will be measured in at the end, we don't see the same gap as noted in (a), but the fidelity is a lot lower than if no mid-circuit measurements were being performed. Middle plot, here we turn off dynamic decoupling on the spectator qubits during mid-circuit measurement and we see a similar pattern to that observed before. The dynamic decoupling, by rotating the qubit during measurement, was allowing fidelity decay to occur regardless of the state of the qubit prior to measurement commencing. Right plot: two different experiments one where the qubit is returned to the $|0\rangle$ before the measurement qubits are measured (blue) and one where it is returned $1\rangle$. The fidelity loss occurs when the qubit relaxes during the measurement of the other qubits. The same effect occurs if we use delays instead of measurements. \textbf{(c)} A dramatic example of the same effect, but here we look at a qubit with very low $T_1$ time. This provides further evidence that what is occurring is a relaxation of the qubit. Right plot, here we randomize the state of the qubit prior to the measurement qubits being measured. This  plot fits the hypothesis that the loss of fidelity observed is a relaxation of the qubit during the time taken for the measurement qubits to be measured.}
    \label{fig:relaxation}
\end{figure*}

The simultaneous RB protocol can be supplemented to additionally characterise mid-circuit measurements; see Fig.~\ref{fig:relaxation}.
While any or all of the qubits can be measured, here we are interested in running logical error correcting codes, providing a partition of qubits into `data' and `ancilla' qubits. 
The qubits that will be measured therefore correspond to the ancilla qubits of the logical code implementation of interest, i.e., those that will be measured to extract the syndrome information. 
The experiment proceeds as before for a number of rounds with single Clifford twirling operations applied each qubit. 
After a number of rounds (in the data presented --- 4), the measurement qubits are returned to their expected final state, and measured, with a $Z$-gate added randomly~\cite{Beale2023}. 
This `round' of four random single Cliffords, measurement and optional Z-gate for the measurement qubits (idle time for the non-measurement qubits) is then repeated a number of times, becoming the `$m$' parameter described in \cref{sec:initialProtocol}. 
The average over several different sequences, under the usual RB assumption, gives an exponential decay~\cite{mclaren2023stochasticerrorsquantuminstruments,zhang2025,Govia2023,hothem2024measuringerrorratesmidcircuit}. 
Here we only look at the results of the final measurement and in particular for each qubit the survival probability for that qubit at the end of each full set of rounds.

From the data gathered by this protocol it is immediately clear that the noise process occurring in the device was not a simple stochastic channel. 
As described in the caption of \cref{fig:relaxation}, further variations of the experiments confirm that both the measurement qubits and the spectator qubits are experiencing `relaxation' error during the course of the mid-circuit measurements, not as a result of the measurements \emph{per se}, but rather simply as a result of the length of time of the measurement.
On this device, reset takes a similar amount of time as measurement.  Therefore, performing a measurement followed by a reset, the total idling error is effectively doubled, with a significant effect on the overall fidelity.  
We believe that this is a combination of amplitude damping and thermal damping noise, based on the measurement time (about 2 \textmu s) compared to the $T_1$ and $T_2$ time (which varies from qubit to qubit, but has a median of 197.36 \textmu s and 118.43 \textmu s respectively. 
The net result on a single qubit twirl for, say, qubit 90 (a proposed data (spectator) qubit) is to reduce the fidelity from  0.998 to 0.983 per measurement cycle.

It was for this reason it became clear that we needed to minimise the number of measurement cycles per cycle of syndrome extraction.

\subsubsection{Temporal consistency}

\begin{figure*}
    
    \includegraphics[width=1\textwidth]{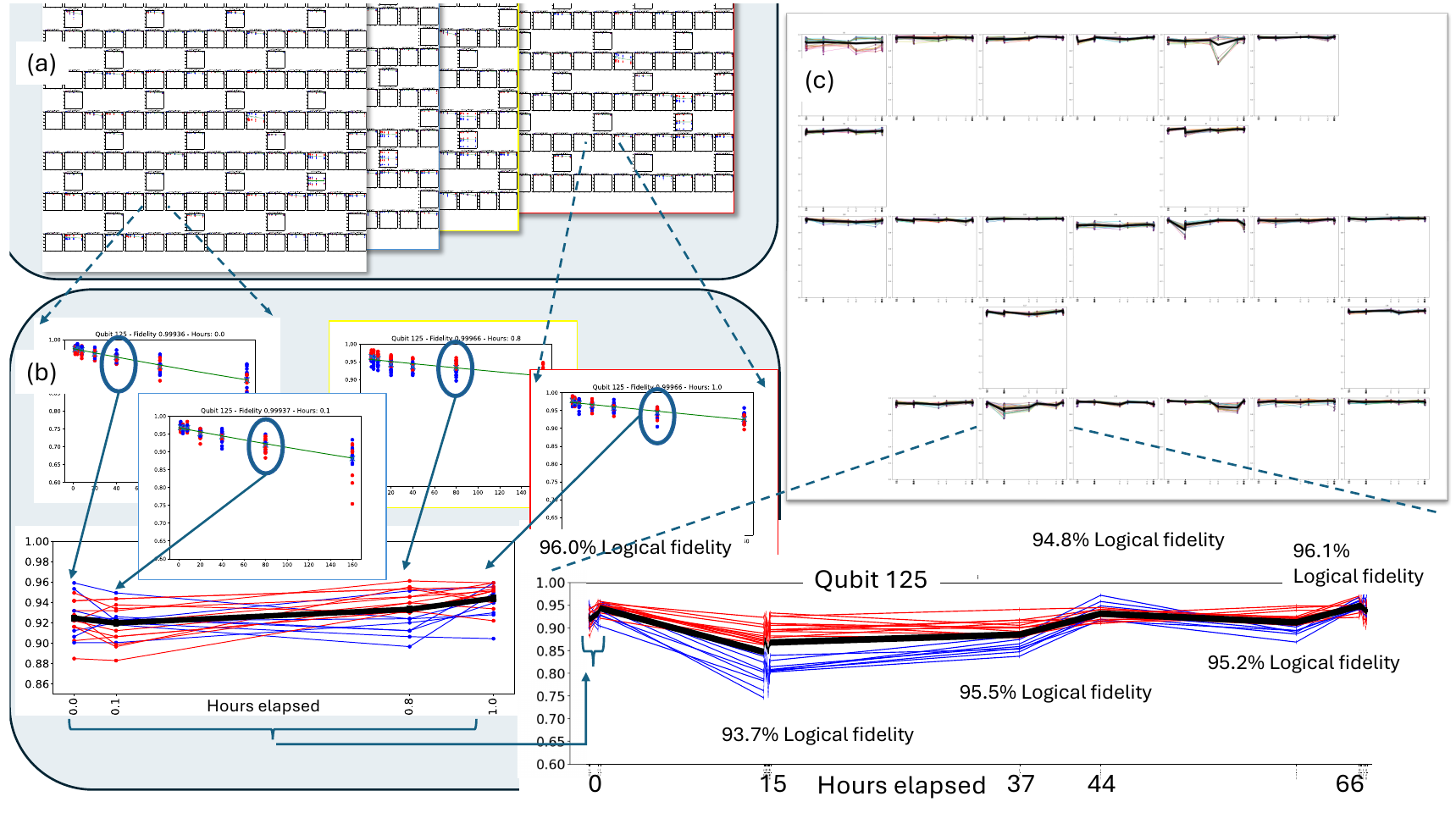}
    \caption{\textbf{Temporal consistency.}  \textbf{(a)} Simultanious randomized benchmarking using full single Clifford twirl circuits are performed on each qubit, interleaved with other experiments (not shown here). The circuits are as described in \cref{sec:initialProtocol}, save that only a single randomized sequence is selected for each qubit. Thereafter, the same sequences are used. \textbf{(b)} For each qubit, we extract the data for a particular length of sequence; here, 80 single qubit gates are chosen. These data are gathered onto the one graph that has as its $x$-axis the time of execution. Each data-point for a sequence (here the average of 512 shots) is connected to the data point for the same sequence, executed at a different time. Each of the lines on the graph should be horizontal, subject to shot noise. The black 'average' line should also be horizontal, with much less shot noise variation. \textbf{(c)} Here we show the data for the qubits in the device constituting a logical qubit. These runs were interleaved with memory experiment runs, which were executed at staggered intervals over a 70 hour period. Inset we extract the graph for one of the qubits (here qubit 125). As can be seen at the 15 hour mark, it appears there was notable change in the ability to correctly read the $|0\rangle$ state (the blue lines). This coincided with a reduction in the logical fidelity of the memory circuits. While we have labeled each point of the this graph with the logical fidelity measured at the time -- at the 44 hour mark the loss of fidelity may be more related to instability in some of the other qubits (top right of top graph). }
    \label{fig:timing}
\end{figure*}

We characterize how consistently the qubit performance parameters are maintained over time, specifically over the medium term, say 30 minutes, and the longer term, say days.  
Again, RB provides an effective and efficient tool for probing this temporal consistency of qubits.  
With simultaneous RB, each individual sequence of random Cliffords will pick up different coherent errors. 
It is only when the sequences are averaged will the results be the equivalent of a depolarizing channel. 
If we were not concerned with confirming the actual fidelity each time we run the protocol then we can select a single randomization for each sequence of the single Clifford gates (or single Pauli gates if we were interested in measuring the non-averaged stochastic Pauli noise). 
Then, unless the device parameters have changed, each subsequent time we run an `experiment' every sequence should give the same result as the previous runs (within `shot' error bounds). 
On the IBM hardware, 30 such sequences are shown in \cref{fig:timing} with 512 shots per sequence, taking well under 1 minute of QPU time (even less if we confine our interest to fewer than 156 qubits). 
Such fast sequences are then straightforward to place in-between other experiments being conducted, allowing us to monitor the quality of the single qubit gates and the quality of the measurements.  This approach should alert us to any changes that might occur in the device during our main experiment. 
This is a relatively simple protocol designed to confirm temporal consistency, see e.g.,\ Ref.~\cite{Proctor2019} for a more complete protocol to allow the tracking of drift.
\vspace{6pt}

\subsubsection{Cross-talk from mid-circuit measurements on syndrome extraction circuits}
\label{sec:mid-circuit}

\begin{figure*}
    
    \includegraphics[width=1\textwidth]{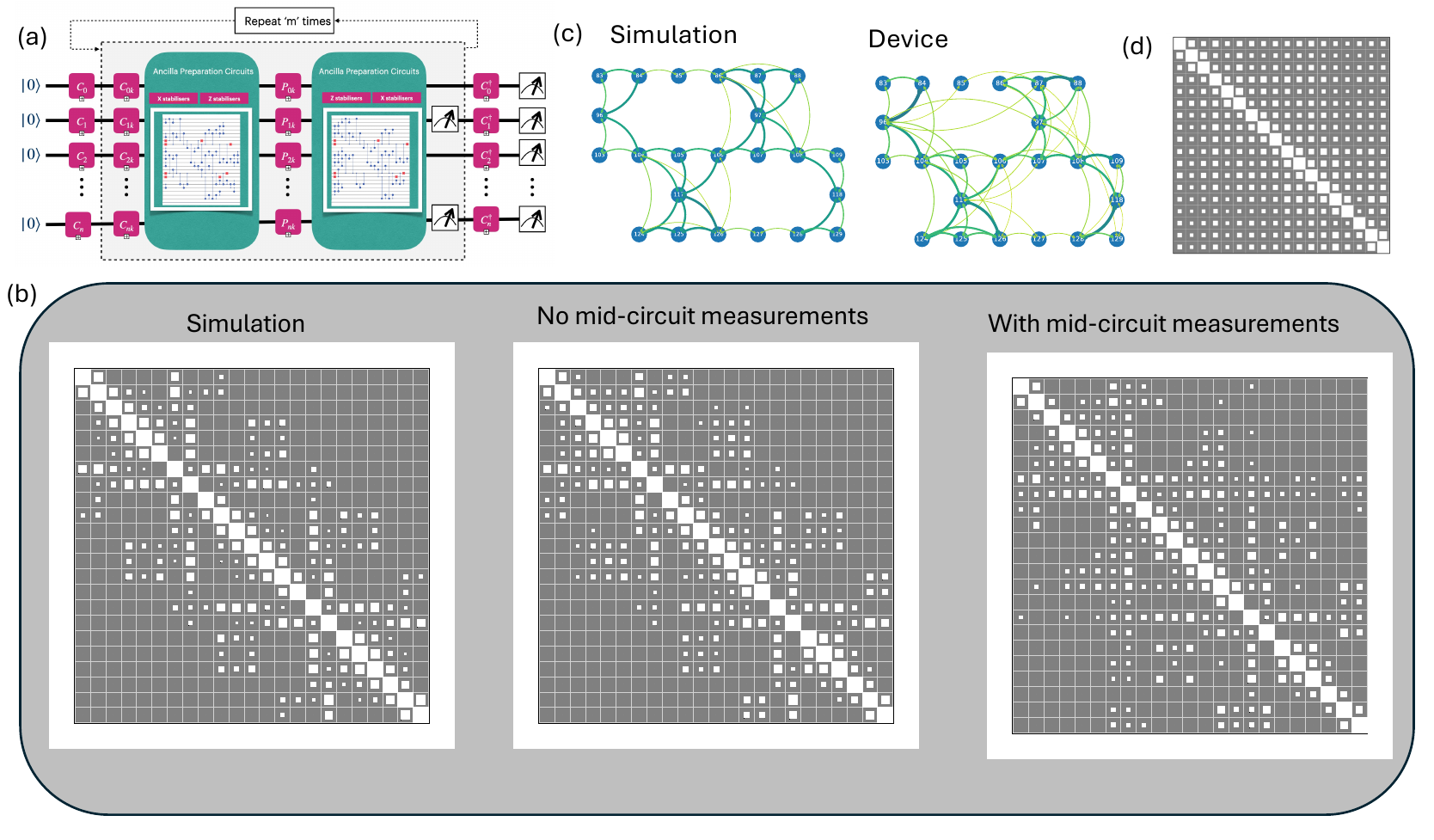}
    \caption{\textbf{Cross-talk from mid-circuit measurements on syndrome extraction circuits.}  \textbf{(a)} Here we adapt the circuit used in Ref.~\cite{Harper2023} for the heavy-hex code with mid-circuit measurements. 
    The channel we measure is a circuit to prepare the $X$ stabilizers, followed by the $Z$ stabilizers, followed by a random Pauli. 
    Then we reverse the process and prepare the $Z$ stabilizers and the $X$ stabilizers. 
    The result is equivalent to an identity circuit, up to a Pauli operator. 
    We can measure the measurement qubits and twirl this block with random single gate Cliffords. 
    This is repeated $m$ times, then a final inverting Clifford is applied. 
    Using this circuit we can extract the fidelity of each qubit in the channel as well as all $k$-body error correlations. 
    For the purposes of this work, we focus on two-body errors. 
    \textbf{(b)} Since the syndrome extraction circuits contain two-qubit entangling gates, errors will spread between the qubits. As this is a fault-tolerant syndrome extraction circuit, the spread of errors is contained if a circuit-level model is being honoured by the device. 
    We can simulate the circuits with Pauli noise and determine the error correlations we would expect to see if there were no additional crosstalk in the device. 
    Left: we plot the expected correlations between qubits in a Hinton diagram. 
    Each row and column represents a qubit. 
    The correlation between qubits is shown by the area of white in the intersecting square, with a full square representing 1. 
    The diagonals are all full white because each qubit is fully correlated with itself. 
    The resulting patterns occur because of the one dimensional projection of the position of the qubits and the symmetry around the diagonal. 
    Middle: the actual crosstalk pattern from the device in the absence of mid-circuit measurements. 
    As can be seen the pattern broadly follows the simulation presented in the left plot, indicating little non-circuit model crosstalk, although some discrepancies can be seen. 
    Right: two qubit correlations where mid-circuit measurements have been added.
    As can be seen there is some additional crosstalk between the qubits, but the expected circuit-level noise model is broadly replicated. 
    \textbf{(c)} The diagrams shown in (b) are compact but lose the information pertaining to the relative position of the qubits. 
    Where we have sparse correlations (as in this case) we can plot the same information using the position diagram of the qubits and connecting qubits with a line if they share information. 
    Here we use mutual information as the discriminating metric and the color and width of the connecting line indicates the strength of the mutual information between the qubits. 
    The actual values of the mutual information will depend on the noise in the system. The notable feature here, however, is the existence of mutual information between qubits that are designed to be quarantined by the operation of the syndrome extraction circuit.
    The most worrying type of crosstalk would be between data qubits that are designed to only experience independent errors.
    Left shows a circuit level simulation and Right the data from the device. 
    As in (b) the broad features of the simulation can be seen, although there is clearly some additional crosstalk between qubits. 
    \textbf{(d)} A Hinton diagram taken from a similar experiment to that shown in (b) Right panel, run on an early Heron r1 device in January 2024. 
    At that time mid-circuit measurement of certain qubits caused considerable crosstalk between many qubits, which was picked up in the experiment. This was an issue that is mainly resolved on the most recent IBM devices (Heron r2).}
    \label{fig:correlations}
\end{figure*}

Here we analyze the effect of mid-circuit measurements on the device, while it is running syndrome extraction circuits.  See Fig.~\ref{fig:correlations}.
Using the techniques in Ref.~\cite{Harper2019b}, the data gathered from the simultaneous RB experiments (\cref{sec:initialProtocol}) can be used to confirm that is no significant crosstalk issues in the device when only single qubit gates, without mid-circuit measurements, and the data from \cref{sec:midSingleProtocol} can be used to check the case with mid-circuit measurements added. In neither case were any significant cross-talk issues noted. The small amount of cross-talk that occurs is consistent with low levels of leakage from the mid-circuit measurement, but this is minor (we estimate at least an order of magnitude smaller) than the idling errors occurring during measurement and classical  measurement errors. 

\acknowledgments
We are grateful for helpful conversations with A. Cross, O. Dial, R. Gupta, S. Merkel, E. Pritchett, N. Sundaresan and M. Takita.
We acknowledge support from the Intelligence Advanced Research Projects Activity (IARPA), under the Entangled Logical Qubits program through Cooperative Agreement Number W911NF-23-2-0223.  The views and conclusions contained in this document are those of the authors and should not be interpreted as representing the official policies, either expressed or implied, of IARPA, the Army Research Office, or the U.S. Government. The U.S. Government is authorized to reproduce and distribute reprints for Government purposes notwithstanding any copyright notation herein. CL acknowledges support from the Engineering and Physical Sciences Research Council [Grant Number EP/S021582/1].

\end{document}